\documentclass[12pt]{article}
\usepackage{ametsoc}
\usepackage{graphics}
\usepackage{epstopdf}

\catcode`\"=\active \let"=\"
\let\3=\ss

\begin{document}

\title{The widest contiguous field of view at Dome C and~Mount~Graham}
\author{Jeff Stoesz  \thanks{\textit{Corresponding author address:}
         Jeff Stoesz, INAF - Osservatorio di Arcetri, Largo Enrico Fermi 5, Firenze,
         FI 50125, Italy. \newline{E-mail: stoesz@arcetri.astro.it}}
         \and Elena Masciadri \and Franck Lascaux \and Susanna Hagelin\\
         	INAF - Osservatorio Astrofisica di Arcetri, Florence, Italy}


\def\glifft{\mbox{{\it GLiFFT }}}
\def\paola{\mbox{{\it PAOLA }}}
\def\pasp{\it PASP}
\def\mnras{\it MNRAS}
\def\nat{}
\newcommand{\bs}[1]{\boldsymbol {#1}}
\providecommand{\as}{^{\prime\prime}}
\providecommand{\am}{^{\prime}}
\def\deg{^{^{\rm o}}}
\def\gsim{ \lower .75ex \hbox{$\sim$} \llap{\raise .27ex \hbox{$>$}} }
\def\lsim{ \lower .75ex \hbox{$\sim$} \llap{\raise .27ex \hbox{$<$}} }
\newcommand{\eqn}[1]{Eqn.(\ref{eq:#1})}
\newcommand{\sect}[1]{\S\ref{sec:#1}}
\newcommand{\fig}[1]{Fig.\ref{fig:#1}}
\newcommand{\tab}[1]{Table~\ref{tab:#1}}
\newcommand{\app}[1]{\S\ref{app:#1}}

\providecommand{\cn}{$C_n^2$~}
\providecommand{\cnh}{$C_n^2(h)$~}
\providecommand{\cni}{${C_n^2}_i(h_i)$~}

\amstitle

\begin{abstract}
The image quality from Ground-Layer Adaptive Optics (GLAO) can be gradually increased with decreased contiguous field of view. This trade-off is dependent on the vertical profile of the optical turbulence (\cn profiles). It is known that the accuracy of the vertical distribution measured by existing \cn profiling techniques is currently quite uncertain for wide field performance predictions 4 to 20 arcminutes. With assumed uncertainties in measurements from Generalized-SCIDAR (GS), SODAR plus MASS we quantify the impact of this uncertainty on the trade-off between field of view and image quality for photometry of science targets at the resolution limit. We use a point spread function (PSF) model defined analytically in the spatial frequency domain to compute the relevant photometry figure of merit at infrared wavelengths. Statistics of this PSF analysis on a database of \cn measurements are presented for Mt. Graham, Arizona and Dome C, Antarctica. This research is part of the activities of ForOT (3D Forecasting of Optical Turbulence above astronomical sites).

\end{abstract}

\section{Introduction}\label{sec:intro}
Characterization of the optical turbulence in the first few kilometres above the telescope is important for predicting the performance of Ground-Layer Adaptive Optics (GLAO) telescopes as a function of field of view diameter.
Systems that have been proposed will correct visible or near-infrared science fields that are typically 4 arcminutes, and potentially up to 20 arcminutes in diameter and contiguous.
There are several measurement techniques being advanced to provide statistics on the vertical distribution of the structure function coefficient \cnh, and in this paper we explore the impact of a potential bias from generalized-SCIDAR and MASS measurements.
The first of two sites we will investigate is a typical mid-latitude observatory site, Mount Graham (32.7 N, 109.87 W, 3200 meters), measured with generalized-SCIDAR.
There are conifer trees at the summit with a height similar to the SCIDAR telescope's primary mirror, about 8 meters above the ground.
The second is Dome C (75.1 S, 123.3 E, 3260 meters), an Antarctic site with MASS and SODAR measurements by \cite{lawrence04} and balloon measurements by \cite{agabi06}.

\newpage
The GLAO PSF figure of merit that is of particular importance to wide field astronomy is radius of 50\% encircled energy, computed at several points in the contiguous field of view and then averaged.
It will be symbolized as $EE50$ here.
$EE50$ is very closely related to the integration time to achieve some signal to noise ratio in background-limited point source photometry in the field \citep{andersen}, a rather common science application for fields of view 4 to 20 arcminutes in diameter.
Roughly,
\begin{equation}
{\rm integration ~time} \propto EE50^2.
\end{equation}
We will compute $EE50$ starting with an analytically defined phase Power Spectral Density (PSD) for anisoplanatism and fitting error using established theory \citep{paola,tokovinin04}.
\tab{pars} lists the model parameters selected here.
Computation from the analytic PSD is a fast method to discover the performance gradient of $EE50(\theta)$, where $\theta$ is the diameter of the field of view.
\begin{table}[h]
\caption{The parameters and implicit assumptions of the GLAO PSF model.\label{tab:pars}}
\begin{tabular}{rl}
\hline \hline
phase PSD & von K{\'a}rm{\'a}n, $L_o=30$ meters \\
telescope diameter & $D=8$ meters \\
Beacons & 4 point sources at range $H=90$ km at zenith \\
Beacons & evenly distributed on a circle of diameter $\theta$ in the field \\
image wavelength & $\lambda=1.25\mu m$ \\
image locations & sampling a square field of view with vertices that intersect the circle \\
Deformable Mirror & cartesian grid of actuators with pitch, $\Delta$ \\
Deformable Mirror & each actuator has a sinc-like influence function \\
Deformable Mirror & conjugated to height = 0 \\
\hline
\end{tabular}
\end{table}

\begin{figure}[h]
\resizebox{!}{6cm}
{
\includegraphics{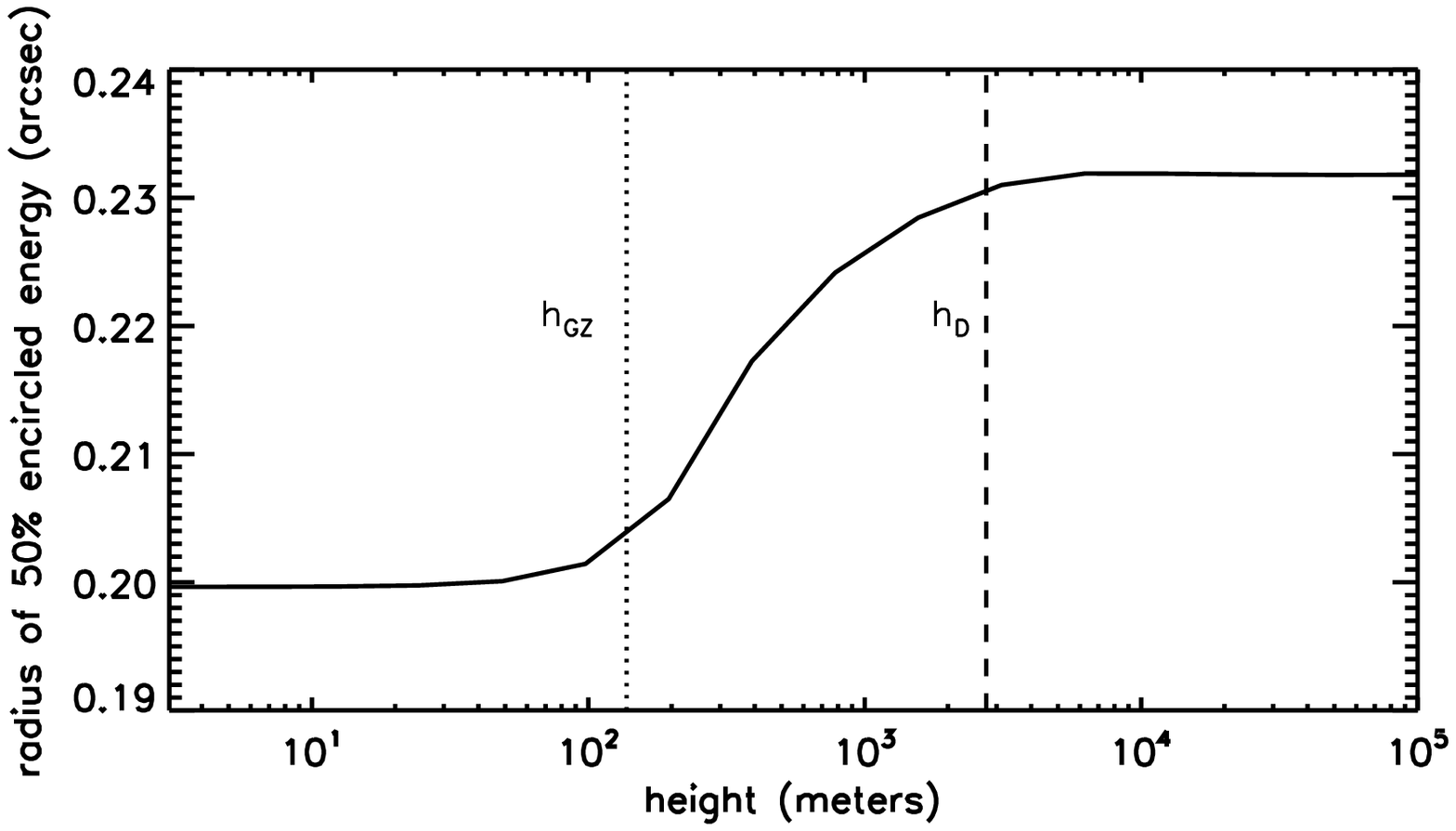}
}
\caption{The gray-zone begins above $h_{GZ}$.}\label{fig:gz}
\end{figure}

The exact range of altitudes in the first few kilometres where bias has greatest impact depends on the basic GLAO system parameters, namely the diameter of the guide star asterism (also $\theta$) whose signal is averaged and the effective pitch that is controlled by the ground conjugated deformable mirror ($\Delta$).
The ratio $h_{GZ}=\Delta/\theta$ defines the altitude below which any contribution to anisoplanatism is negligible.
The term gray-zone (GZ) was coined \citep{tokovinin04} to identify the altitudes above $h_{GZ}$, where the contribution to anisoplanatism is not negligible (also known as partially corrected zone).\footnote{Looking at the approximate error transfer function in equation (8) of \cite{tokovinin04} one can see why this is the case.}
\fig{gz} helps illustrate this in terms of performance in the focal plane.
The plot shows the $EE50$ figure of merit as a function of the height of one layer of turbulence added to a typical, smooth profile.
The layer contains half of the total turbulence strength of the smooth profile.
\fig{gz} shows that the largest performance gradient is at altitudes just above $h_{GZ}$.
The gradient vanishes above $h_D=D/\theta$, where $D$ is the telescope diameter.
In the following sections we will re-compute $EE50(\theta)$ with estimated bias in the proportion of turbulence attributed to heights above or below $h_{GZ}$.

\section{Mount Graham and Dome C profile monitoring data}

The Mt. Graham G-SCIDAR measurements include 851 in High Vertical Resolution (HVR) mode and 9911 in regular mode, both have been reduced to discretized turbulence strength $J_i$ at height $h_i$. 
These were computed from the normalized covariance function of the irradiance fluctuations \citep[see][]{egnerHVR06,egner07} which are proportional to $J_i$, which are in turn related to \cnh by
\begin{equation}
J_i = \int_{{h_b}_i}^{{h_b}_{i+1}} dh~ C_n^2(h). \label{eq:ji}
\end{equation}
The intrinsic vertical resolution of SCIDAR is roughly given by
\begin{equation}
\frac{0.78}{\rho}\sqrt{\lambda |h+h_{gs}|} \label{eq:vres}
\end{equation}
where $\rho$ is the binary separation ($35\as$), $\lambda$ is the wavelength of the scintillation signal ($0.5\mu m$), and $h_{gs}$ is the conjugation height of the generalized SCIDAR analysis plane (about $-3500m$).
The regular mode resolution will represent free-atmosphere, above 1000 meters.
The current HVR data set samples the scale height of the boundary-layer and provides data up to 1000 meters altitude.
In a subsequent section we will describe how the ground-layer and free-atmosphere are reduced to form a composite statistical model.

For Dome C we will use 1701 MASS+SODAR profile monitoring measurements at Dome C by \citet{lawrence04} during the Antarctic winter of 2004.
These data sample only two grid points between 30 and 1000 meters and do not sample any turbulence below 30 meters.
However, there exist balloon-borne micro-thermal measurements \citep{agabi06} that give us an estimate of the scale height and total strength of the ground-layer, and with this information we model the statistics of eight grid points from a height of zero to 200 meters.
The turbulence measurements recorded by SODAR in the \cite{lawrence04} data we appropriate to a slab concentrated at 250 meters between the modelled ground layer and the lowest MASS measurement at 500 meters.

For the Dome C altitudes from zero to 200 meters we define the following exponential model to 
\begin{equation}
C_n^2(h) = Ae^{(-h/h_A)}.
\end{equation}
Using \eqn{ji} it follows that
\begin{equation}
J_i = -Ah_A\left(e^{(-{h_b}_{i+1}/h_A)}-e^{(-{h_b}_i/h_A)}\right). \label{eq:dcji}
\end{equation}
We will choose the boundaries ${h_b}_i$ in \sect{resample}.
Using a average, weighted by \cnh
\begin{eqnarray}
h_i  = & \frac{\int_{{h_b}_i}^{{h_b}_{i+1}}dh~ C_n^2(h) ~h}{ \int_{{h_b}_i}^{{h_b}_{i+1}}dh~ C_n^2(h)}. \nonumber \\
 = & \frac{-Ah_A\left[({h_b}_{i+1}+h_A)e^{(-{h_b}_{i+1}/h_A)}-({h_b}_{i}+h_A)e^{(-{h_b}_i/h_A)}\right]}{J_i}. \label{eq:dchi}
\end{eqnarray}
It has been observed with balloon measurements at Cerro Pachon \citep{tokomodels} that the strength of ground-layer is governed primarily by the scale height.
In our model we will make the scale height dictate the strength exclusively.
A lognormal distribution of values of the scale height, $h_A$, while $A=740.\times 10^{-16}$ and is fixed, will give a lognormal distribution in seeing.

\begin{figure}[h]
\resizebox{!}{6cm}
{
\includegraphics{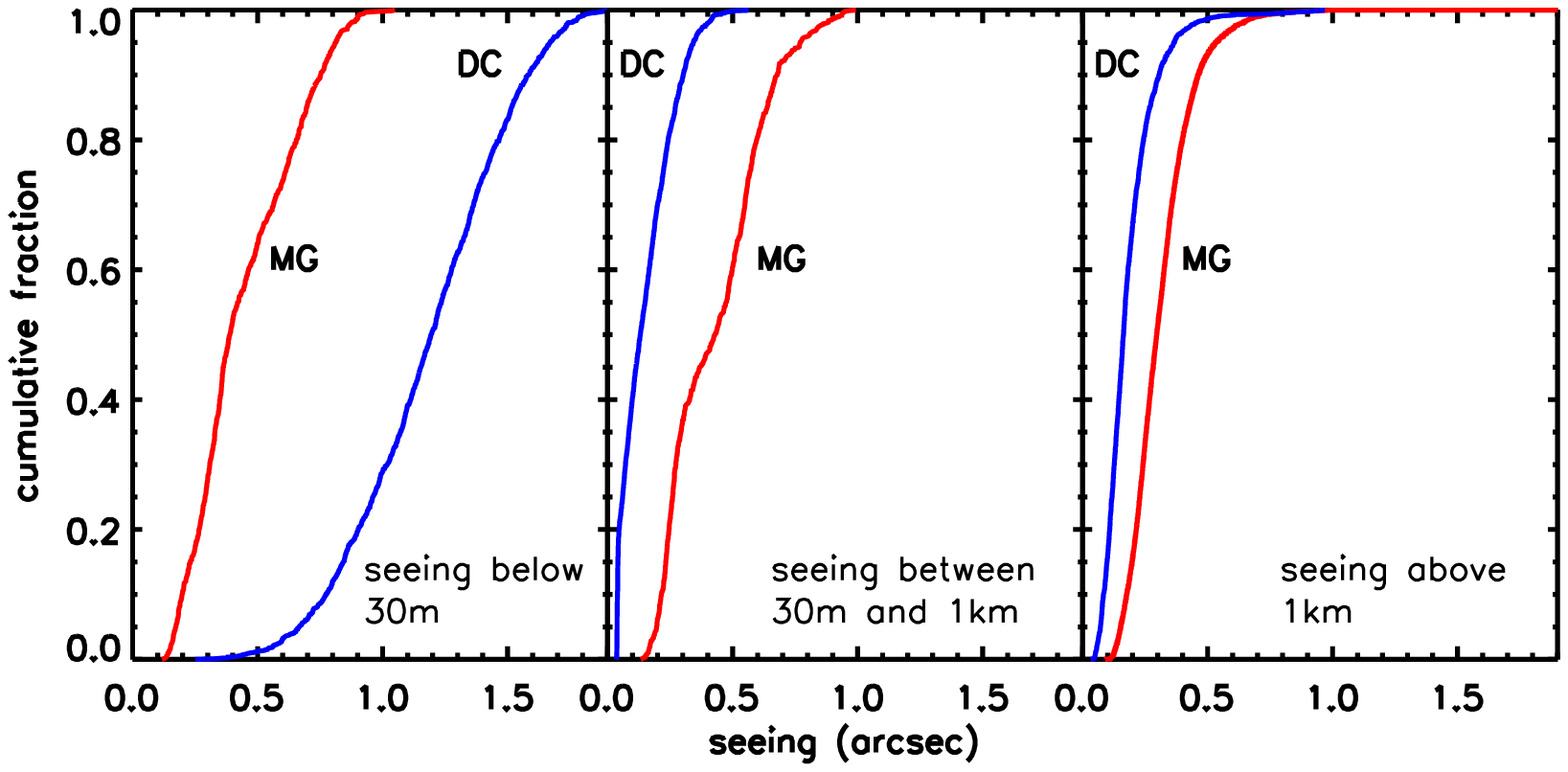}
}
\caption{Comparison of the Dome C (DC) and Mount Graham (MG) turbulence profile data used here.}\label{fig:profiles}
\end{figure}

\clearpage
The Mt. Graham (MG) scenario has weaker overall seeing (median 0.74 arcseconds) than Dome C (DC, median 1.2 arcseconds).
To illustrate the differences in the vertical distributions for these two sites we reduce the data to cumulative histograms of seeing in three slabs, shown in \fig{profiles}.
The Dome C free atmosphere (right panel) and even upper ground-layer slab (middle)are quite calm.
Though the left and middle panels of \fig{profiles} are not proof, the scale height of the MG turbulence is resolved by the HV-GS technique in another analysis \citep{egnerHVR06} to be between 100 to 250 meters.
The DC scenario clearly has most turbulence concentrated between the telescope and 30 meters range (left panel \fig{profiles}).

\section{Reduction to composite profiles}\label{sec:resample}
Since the measurements of the ground-layer and free-atmosphere at these sites is not simultaneous, we must create composite profiles that would closely reproduce the PSF statistics as though we had computed them on a full set of $J_i(h_i)$ data, uninterrupted in $h$ and sampled at the same time.
To do this we sort and combine the profiles of as described in \cite{tokomodels} using the assumption of uncorrelated ground-layer and free-atmosphere seeing. 
We will briefly re-describe the process here in the context of our data.

The Mt. Graham HVR will provide the ground-layer below 1000 meters and the regular SCIDAR measurements will provide the free-atmosphere above 1000 meters.
Three groups of profiles in the ground-layer are identified using the sum of $J_i$.
The first group are those profiles within $5\%$ of the $25^{th}$ percentile are combined in a simple average for $J_i$.
We call them the ``good'' case.
The $50^{th}$ and $75^{th}$ percentile profiles area combined similarly and called ``typical'' and ``bad''.
In each group the grid of $h_i$ is identical and hence remains unchanged by the combining process.
The same process is done for the free-atmosphere.
The result is a reduction to three ground layer profiles and three free-atmosphere profiles, which together have nine permutations for composite profiles that can reproduce the PSF statistics as though we had computed them on all of the $J_i(h_i)$ data.

For Dome C we sort and combine the MASS+SODAR profile monitoring measurements of the free-atmosphere above 200 meters in the same way we described for Mt. Graham.
The ground-layer model does not need to be sorted; the choice of three scale heights $h_A=[14,9,22]$ meters provide the median, first and last quartile of the integrated ground-layer.

\section{Resampling the Composite Profiles} \label{sec:resample}
In all cases the shape of the composite profiles, whether averaged over time or defined by a function is smooth and well sampled by the grid of $J_i(h_i)$ defined so far.
Hence, we are permitted to resample the the $J_i(h_i)$ grid for the GLAO PSF model, which is affected by the density of points in the gray-zone.
We increase the number of grid points in the gray-zone until the PSF figure of merit has reached an asymptote. 
This is trivial for the ground-layer of Dome C, we can define the $h_b$ grid and then re-compute $J_i(h_i)$ with \eqn{dcji} and \eqn{dchi}.
For the measurements of Mount Graham and the free-atmosphere of Dome C we divide several measured $J_i(h_i)$ grid into more numerous $J_j(h_j)$ using linear interpolation of the original discretized \cnh data.

\section{Predicted GLAO performance gradient} \label{results}

\begin{figure}[h]
\resizebox{\hsize}{!}
{
\includegraphics{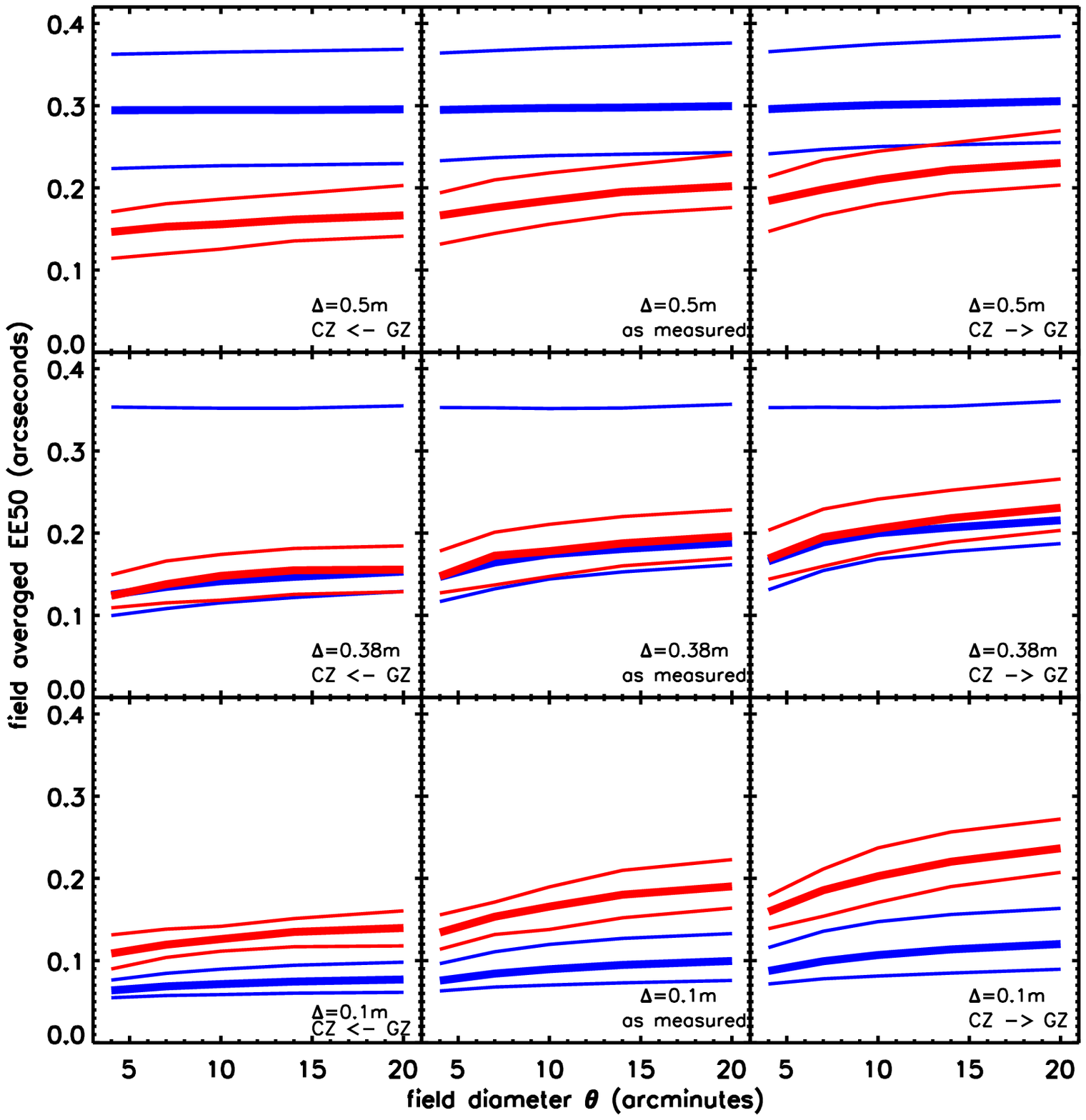}
}
\caption{The field averaged radius of 50\% encircled energy on PSFs at $1.25\mu m$, plotted as a function of the GLAO field of view.}\label{fig:ee50}
\end{figure}

The reduced composite \cnh profiles for each site are input for the computation of field averaged radius of 50\% encircled energy of PSFs at a wavelength of $1.25\mu m$, outlined in \sect{intro}, and symbolized $EE50$.
The aim is to asses the impact on GLAO performance by potential biases in the measured vertical distribution of the turbulence strength.
We have selected the performance $EE50(\theta)$ metric to do this.
\fig{ee50} is a 3x3 multi-panel plot showing $EE50(\theta)$ at Mt. Graham (red) and Dome C (blue).
The thicker lines are the median values while the thinner ones are the first and last quartiles of the ordinate.

Let us first consider the central column of plots to identify the fundamental differences between weak and strong free-atmosphere sites.
In the upper one we see the Mt. Graham (red) $EE50$ gracefully increasing with $\theta$, as the bottom of the gray-zone (\sect{intro}) reaches into the boundary-layer turbulence 100 to 250 meters thick.
For this top middle panel the actuator pitch of the DM was 0.5 meters and the Dome C scenario only very weakly affected by anisoplanatism, a consequence of an inadequate number of actuators for that site.
In the central panel the pitch is 0.38 meters, which improves correction at Mt. Graham slightly in all conditions, and greatly improves Dome C for median or better conditions.
The median and first quartile $EE50(\theta)$ curves of Dome C and Mt. Graham have similar shape because the ground-layer profiles at Mt. Graham have similar exponential shape.
The bottom plot shows the potential gain for Dome C when the wavefront is controlled to a pitch of 0.1 meters.
In the central column of plots, the important distinction between the two sites is that Dome C is always under-actuated with $\Delta=0.5$ and sometimes near the diffraction-limited $EE50$ with $\Delta=0.1$.
Mt. Graham on the other hand has more high altitude turbulence and is always limited by anisoplanatism for these $\Delta$.

Next, consider the columns of panels to the left and right of \fig{ee50} showing uncertainties pertinent to field of view trade-offs in GLAO telescope design.
As indicated in figure 4 in \cite{some_SCIDAR_paper} both MASS and SCIDAR measurements are believed to produce faithful total integrals of turbulence, however, the vertical distribution may be biased.
The left column of plots in \fig{ee50} were computed from the $J_i(h_i)$ times 0.5 in the domain $h_{gz}<h_i<6km$, the balance was conserved by putting turbulence in the lowest layer, below $h_{GZ}$.
Likewise the the right column of plots is $J_i(h_i)$ times 1.5 in the domain $h_{gz}<h_i<6km$, with the balance conserved by removing turbulence from the lowest layer.
The change from the central column of plots to the left or the right is the slope of the curves, germane to designing a field of view trade-off.
The performance of a wide field survey can be expressed using the number of square arcminutes of sky that can be imaged to some limiting magnitude per unit time.
For an theoretical seeing-limited telescope this is of course proportional to $\theta^2$.
For a GLAO telescope with field of view $\theta$ it will be roughly proportional to $(\theta/EE50(\theta))^2$.
$EE50(\theta)$ in the middle row of \fig{ee50} ($\Delta=0.38$ meters) the slope of the median Mt. Graham $EE50(\theta)$ in the domain $10<\theta<20$ arcminutes is about $45\%$ less or more in the left or right panels.
It is about $\mp 15\%$ for Dome C.
In terms of $integration~time(\theta)\propto EE50(\theta)^2$ in the domain $10<\theta<20$ we find the slope is $\pm 60\%$ for Mt. Graham, $\pm 30\%$ for Dome C.
In other words, at a mid-latitude site similar to Mt. Graham, the predicted survey coverage of the GLAO telescope could potentially be wrong by as much as 60\%.

\clearpage
\section{Summary}
The GLAO telescope scenario simulated here is a common design for wide field science demanding a contiguous field.
The estimate of 50\% uncertainty in the proportion of turbulence strength between the  the corrected-zone and the gray-zone (in the first 6 $km$) is based on a comparison between MASS and SCIDAR and here we calculate an uncertainty of 60\% in the slope function $EE50(\theta)$.
Dome C is truly a unique site, and more immune to the 50\% uncertainty.
However, if the true uncertainly is not simply multiplicative the uncertainty propagated to $EE50(\theta)$ for Dome C might be similar to that of Mt. Graham.

\emph{Acknowledgements.}

We would like to thank the authors of \cite{lawrence04} for providing their SODAR+MASS data.  This work has been funded by the Marie Curie Excellence Grant (ForOT)-MEXT-CT-2005-023878.

{\bibliography{astro_and_ao_new}
\bibliographystyle{ametsoc.bst} }

\end{document}